\documentclass[namedreferences]{SolarPhysics}
\usepackage[optionalrh]{spr-sola-addons} 
\usepackage{graphicx}        
\usepackage{color}           
\usepackage{url}             

\begin{document}

\begin{article}

\begin{opening}

\title{Subsurface Flows in and Around Active Regions with Rotating and Non-rotating Sunspots \\
}

\author{K.~\surname{Jain}\sep
R.W.~\surname{Komm}\sep
I.~\surname{Gonz{\'a}lez~Hern{\'a}ndez}\sep
S.C.~\surname{Tripathy}\sep
F.~\surname{Hill}
       }
\runningauthor{Jain et al.}
\runningtitle{Subsurface Flows and Sunspots}

   \institute{$^{1}$ National Solar Observatory, Tucson, AZ 85719, USA\\
                     email: \url{kjain@nso.edu} email: \url{rkomm@nso.edu} 
email: \url{irenegh@nso.edu} email: \url{stripathy@nso.edu} email: \url{fhill@nso.edu} \\
             }

\begin{abstract}
The temporal variation of the horizontal velocity  in subsurface layers
beneath three different types of active regions is studied using the technique of ring diagrams. 
In this study, we select active regions (ARs) 10923, 10930, 10935 from three consecutive Carrington rotations: 
AR 10930 contains a fast-rotating sunspot in a strong emerging active region 
while other two have non-rotating sunspots with emerging flux in AR 10923 and decaying flux in AR 10935.
The depth range covered is  from the surface to about 12 Mm.
In order to minimize the influence of systematic effects, the selection of active and quiet regions 
is made so that these were observed at the same heliographic locations on the solar disk. 
We find a significant variation in both components of the horizontal velocity
in active regions as compared to quiet regions. The magnitude is higher 
in emerging-flux regions than in the decaying-flux region, in agreement  
with earlier findings. Further,  we clearly see a significant temporal variation in depth profiles
 of both zonal and meridional flow components  in AR 10930, with 
the variation in the zonal component being more pronounced.  We also notice a significant 
influence of the plasma motion in areas closest to the rotating sunspot in AR 10930 while
areas surrounding the non-rotating sunspots in all three cases are least affected by the presence of
the active region in their neighborhood.

\end{abstract}
\keywords{Helioseismology, Observations - Interior, Convection Zone - Velocity Fields, Interior - 
Sunspots, Velocity}
\end{opening}

\section{Introduction}
Many intriguing studies related to the Sun in recent years have been focused on
 deriving detailed information about active regions (ARs). Basically, these are the 
regions with high magnetic field concentration that
may consist of one sunspot or an ensemble of several sunspots. The sunspots 
are seen to move across the disk for a long time and have been used to estimate differential 
rotation  in the convection zone ({\it e.g.} \opencite{jj05}; \opencite{dh90}; and references therein). 
However, in some cases, these sunspots are also seen to rotate around their umbral centers 
or other sunspots within the same active region. The sunspots in this category are generally
 known as ``rotating sunspots'' (\opencite{evershed09}; \opencite{knovska75}; \opencite{brown03}; \opencite{soon09}). Studies suggest 
that there is no preferential direction for their rotation,  some rotate in clockwise direction 
while others in anti-clockwise direction. Furthermore, this kind of rotation may lead to the 
buildup of energy, which might be later released by a flare \cite{stenflo69}. 
The origin of such rotational motion is thought to arise from the shear and 
twist in magnetic-field lines or {\it vice versa}. It is also suggested that the magnetic 
twist may result from the disturbance in large-scale flows in the solar convection zone and 
the photosphere or sub-photospheric layers \cite{lopez03}.

In order to explore the dynamics beneath the solar surface, 
one may use the techniques of local helioseismology, which are capable of 
probing the solar interior in three dimensions 
 (\opencite{antia07}; \opencite{gizon10}; and references therein).  These techniques
allow us to infer flows in different layers from the surface to several Mm in depth. 
For example, \inlinecite{zhao03} applied the method of time--distance analysis 
to Michelson Doppler Imager (MDI) Dopplergrams to infer sub-photospheric vertical flows
in a fast rotating sunspot region (AR 9114) and found evidence for two opposite 
vertical flows in the depth range of 0--12 Mm. Strong converging flows were found in
the upper layers (0--3 Mm) while divergent flows were observed in deeper layers 
( 9--12 Mm). Another technique, the ring-diagram method, has also been 
used to study flows in sub-surface layers in synoptic maps and active regions 
(\opencite{haber02}; \opencite{irene06}; \opencite{komm04}, \citeyear{komm09}). To verify the reliability of the results 
obtained with different techniques, \inlinecite{brad04} carried out a detailed 
comparison of horizontal flows using data from two years of the  MDI Dynamics Program 
and found a good correlation between flows obtained with 
ring--diagram and time--distance methods. 

In this article, we apply the ring-diagram technique to investigate the horizontal velocity in 
sub-photospheric layers beneath three active regions. Previous studies using this technique for active 
regions are mostly limited to average behavior summed over many  regions irrespective 
of their dynamical characteristics ({\it e.g.} \opencite{komm11}; and references therein). Thus, the temporal
 behavior of sub-surface flows in individual active regions is not fully addressed.
Here, we categorize active regions on the basis of their characteristics and examine 
sub-surface horizontal velocity in and around those regions. We further investigate how velocity components
beneath a rotating sunspot differ from that in a non-rotating sunspot. We infer depth profiles
in the outer 2\%  of the solar interior by inverting velocities obtained from 
ring diagrams, which is briefly described in Section 2. The selection of data
is discussed in Section 3. Finally, results are presented in Section 4 and summarized in Section 5.

\section{Technique}

 In the ring-diagram technique, medium to high-degree acoustic modes are used to infer the characteristics 
of the propagating waves in localized areas of the solar surface \cite{hill88}. This method has 
been  extensively used to study subsurface properties of both active and quiet regions 
(\opencite{rajaguru01}; \opencite{basu04}; \opencite{jain08}; \opencite{rick08};
 \opencite{sushant08}; \opencite{komm08}, \citeyear{komm09}; \opencite{komm11}, \opencite{baldner11}). In this analysis,
we select an area of $\approx$11$^{\mathrm{o}}\times$11$^{\mathrm{o}}$ centered on the active region 
on each image and tracked for 1680 minutes (referred to as one ring day in this article) using the 
surface rotation rate \cite{snod84}.
 Each tracked area is apodized with a circular function and then a three-dimensional FFT is applied on 
both spatial and temporal direction to obtain a three-dimensional power spectrum. We fit the 
corresponding power spectrum using a Lorenzian profile model \cite{haber00},
\begin{eqnarray}
P (k_x, k_y, \omega) & = & {A \over (\omega - \omega_0 + k_xU_x+k_yU_y)^2+\Gamma^2} \nonumber \\
&& + {b \over k^3}
\end{eqnarray}
where $P$ is the oscillation power for a wave with a temporal frequency ($\omega$) and
the total wave number $k^2=k_x^2+k_y^2$. There are six parameters to be fitted:
two Doppler shifts ($k_xU_x$ and $k_yU_y$) for waves propagating in the orthogonal 
zonal and meridional directions, the background power ($b$), the mode
central frequency ($\omega_0$), the mode width ($\Gamma$), and the amplitude ($A$).
Finally, the fitted velocities ($U_x$ and $U_y$) are inverted using regularized least square 
(RLS) method to estimate depth dependence of various components of the horizontal velocity. 

\begin{figure}   
   \centerline{
\includegraphics[width=40mm]{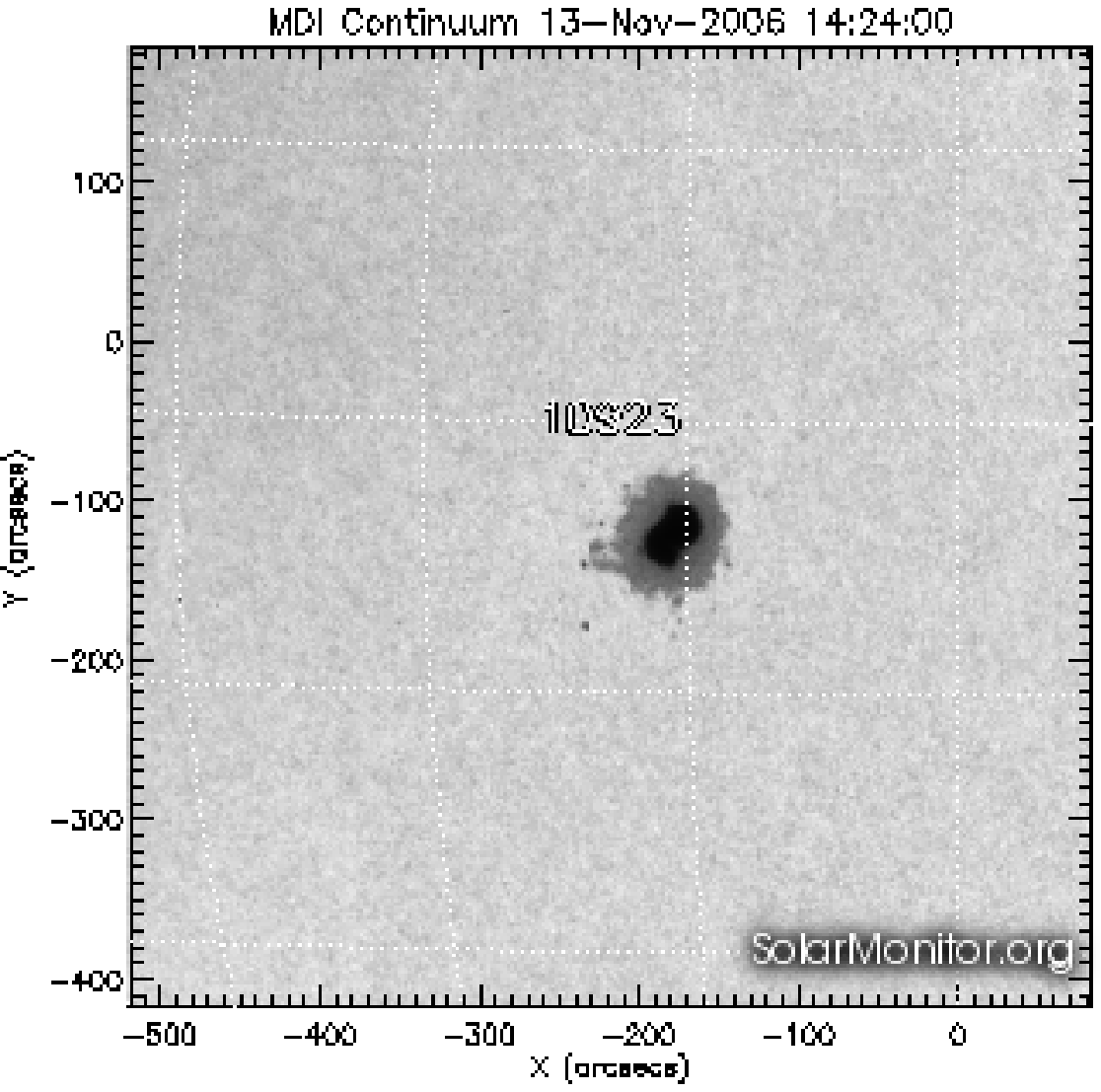}\\
\includegraphics[width=40mm]{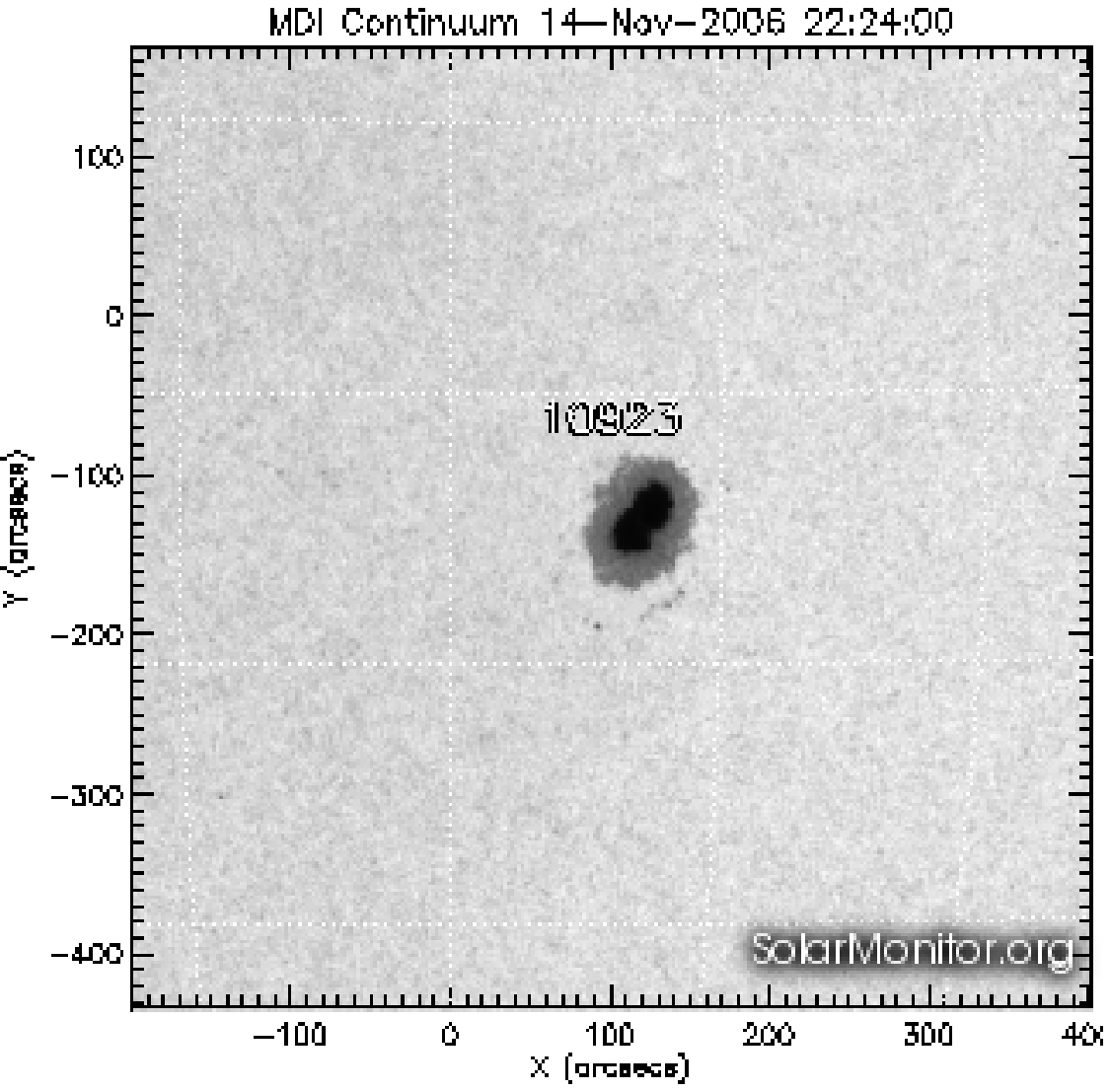}\\
\includegraphics[width=40mm]{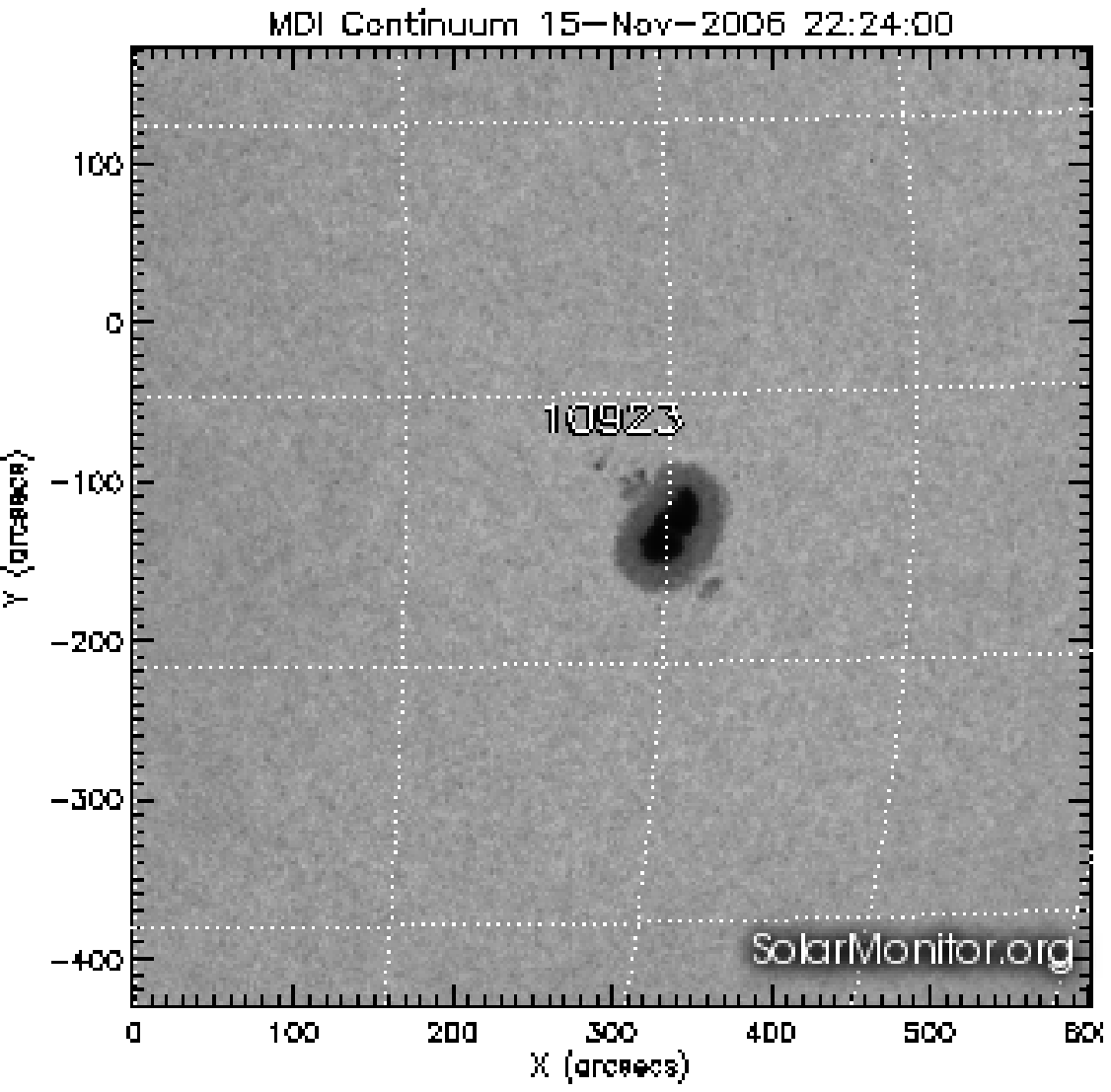}
              }

   \centerline{
\includegraphics[width=40mm]{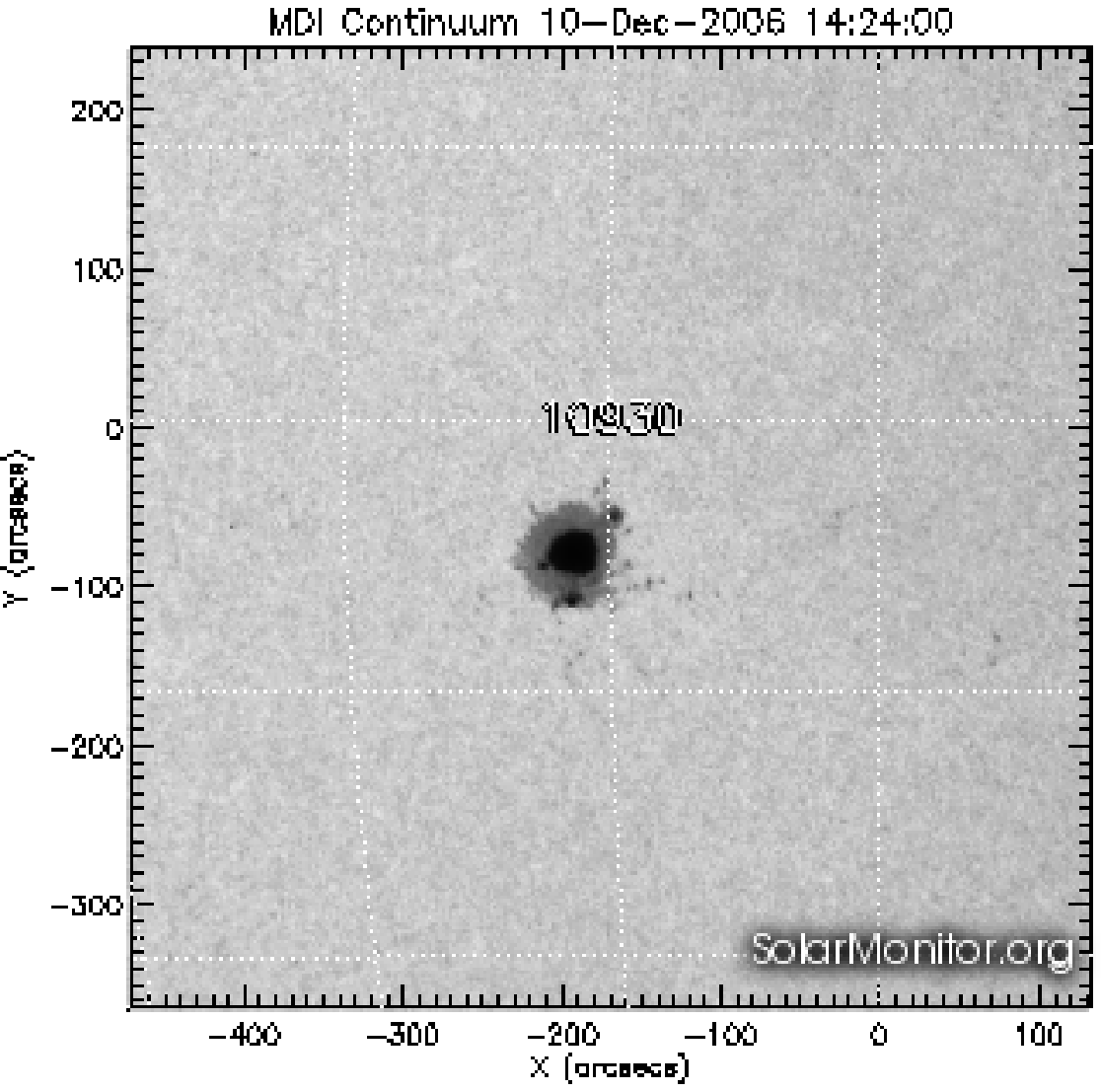}\\
\includegraphics[width=40mm]{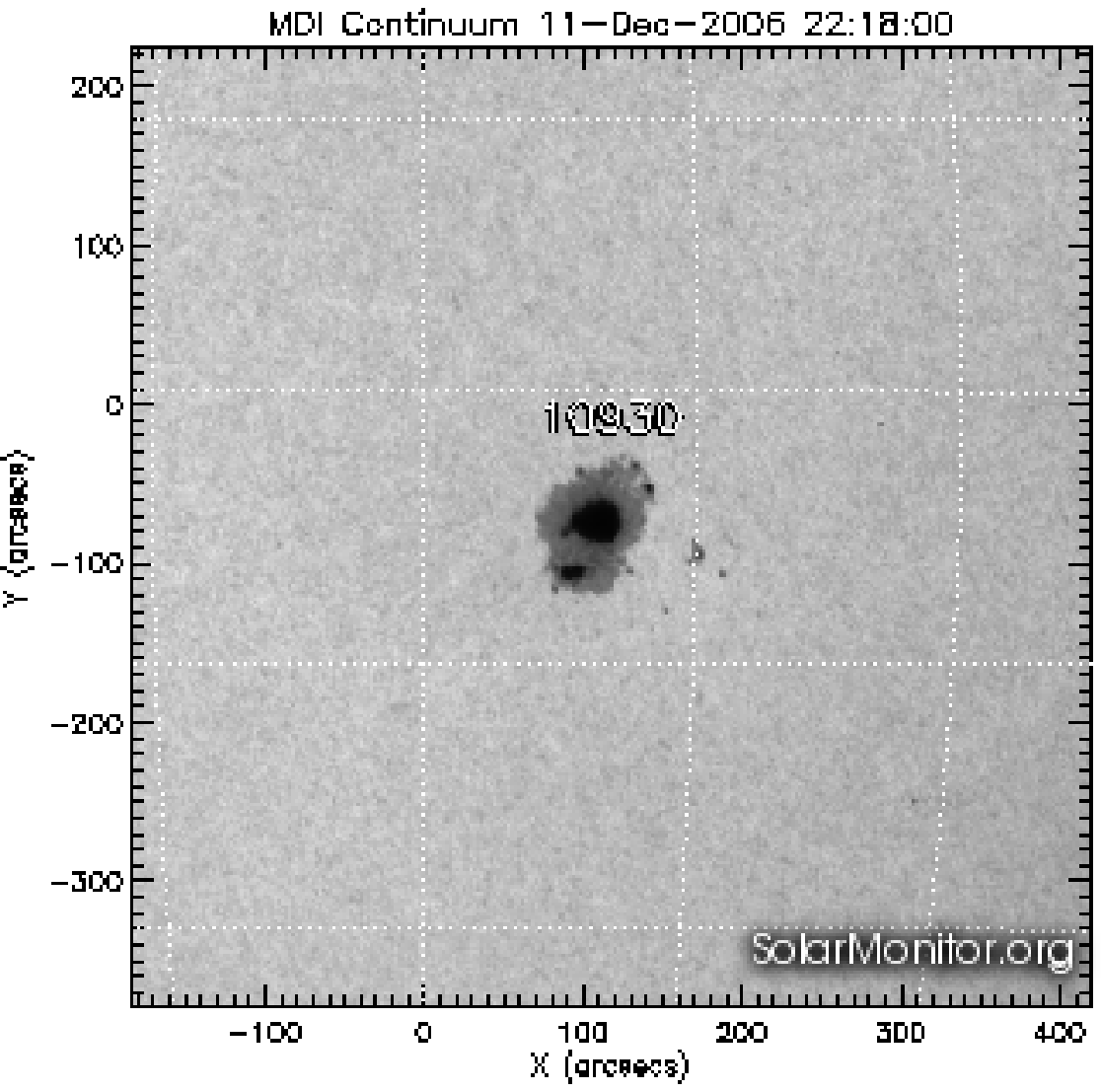}\\
\includegraphics[width=40mm]{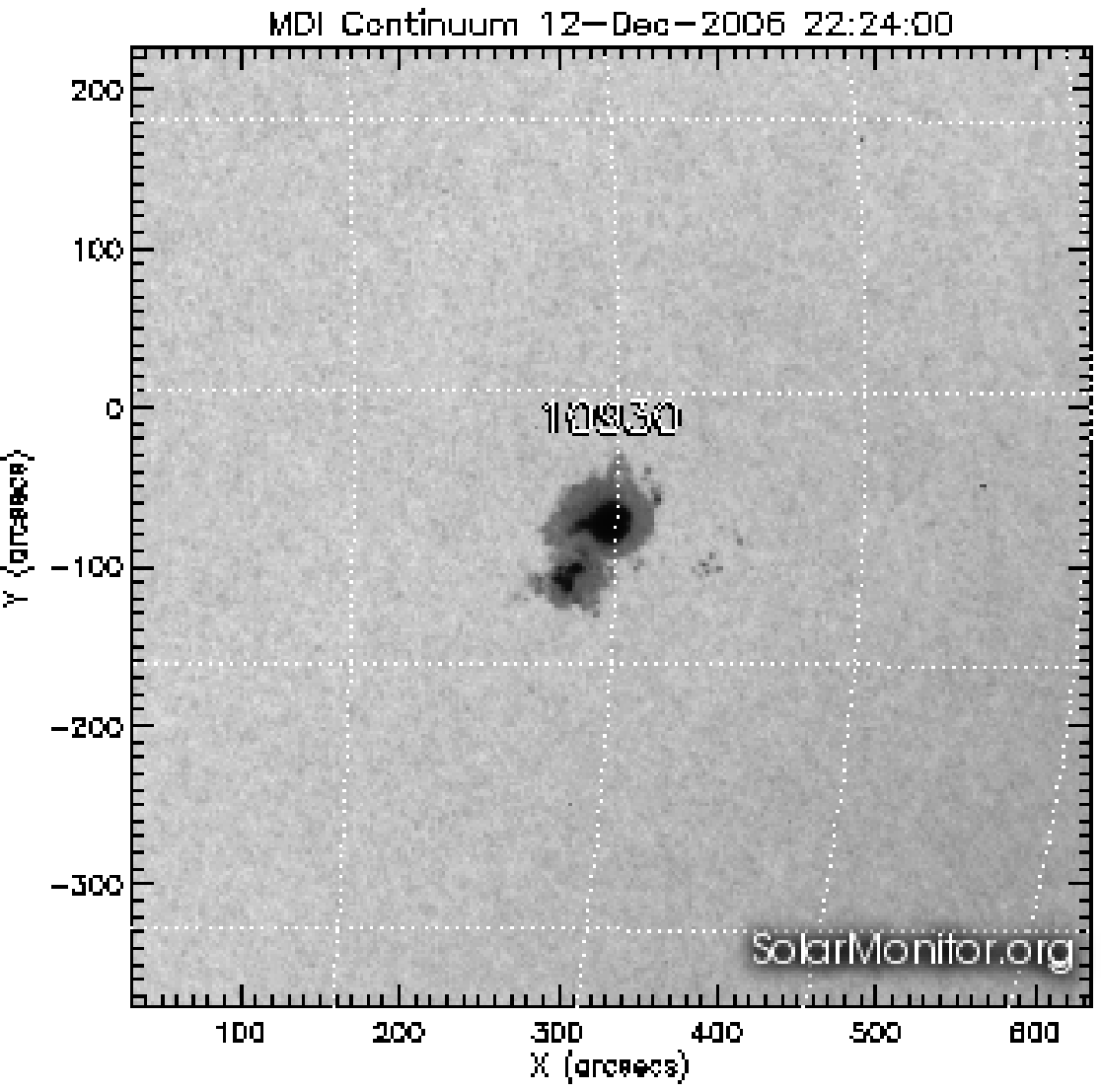}
              }
   \centerline{
\includegraphics[width=40mm]{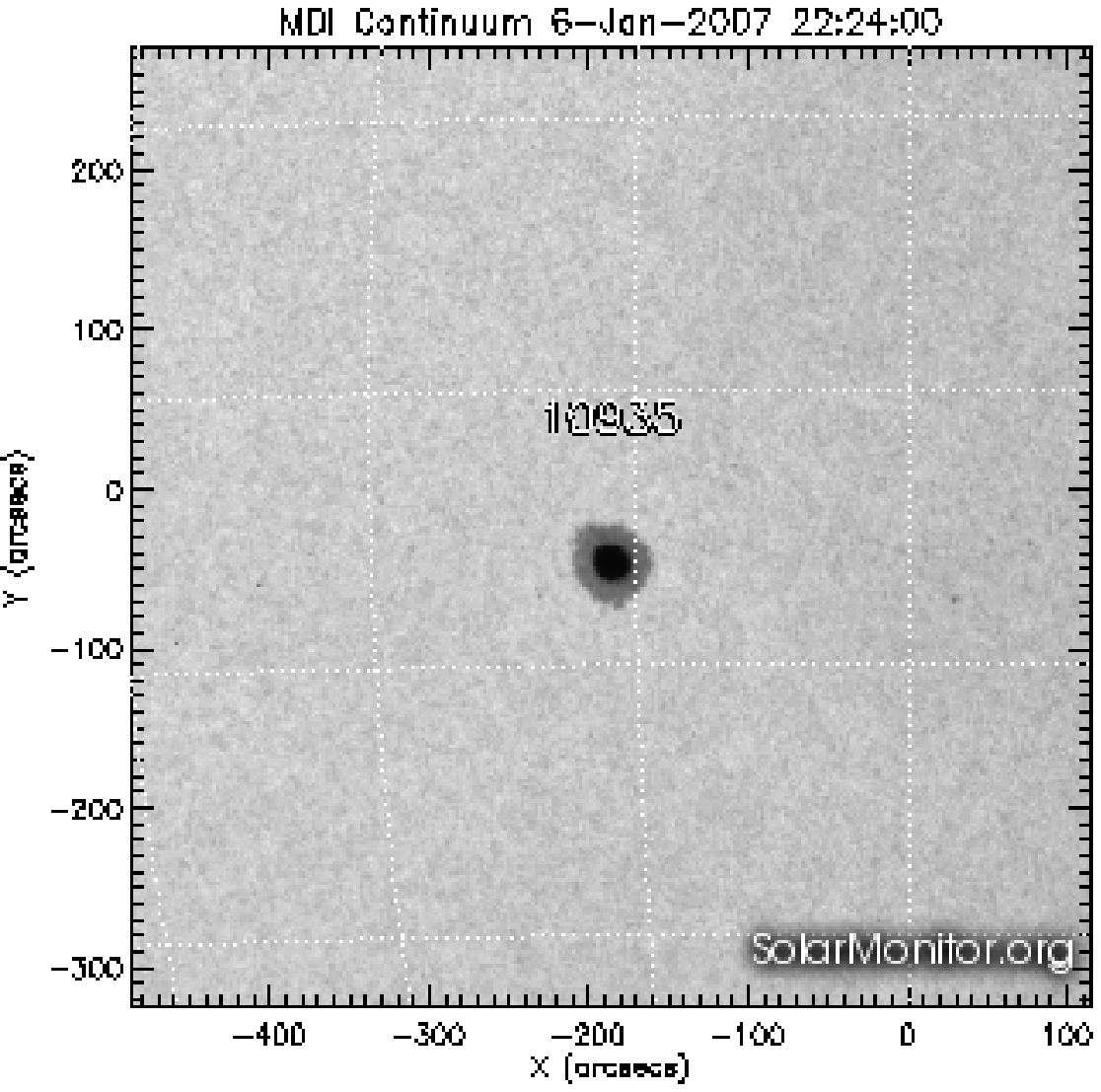}\\
\includegraphics[width=40mm]{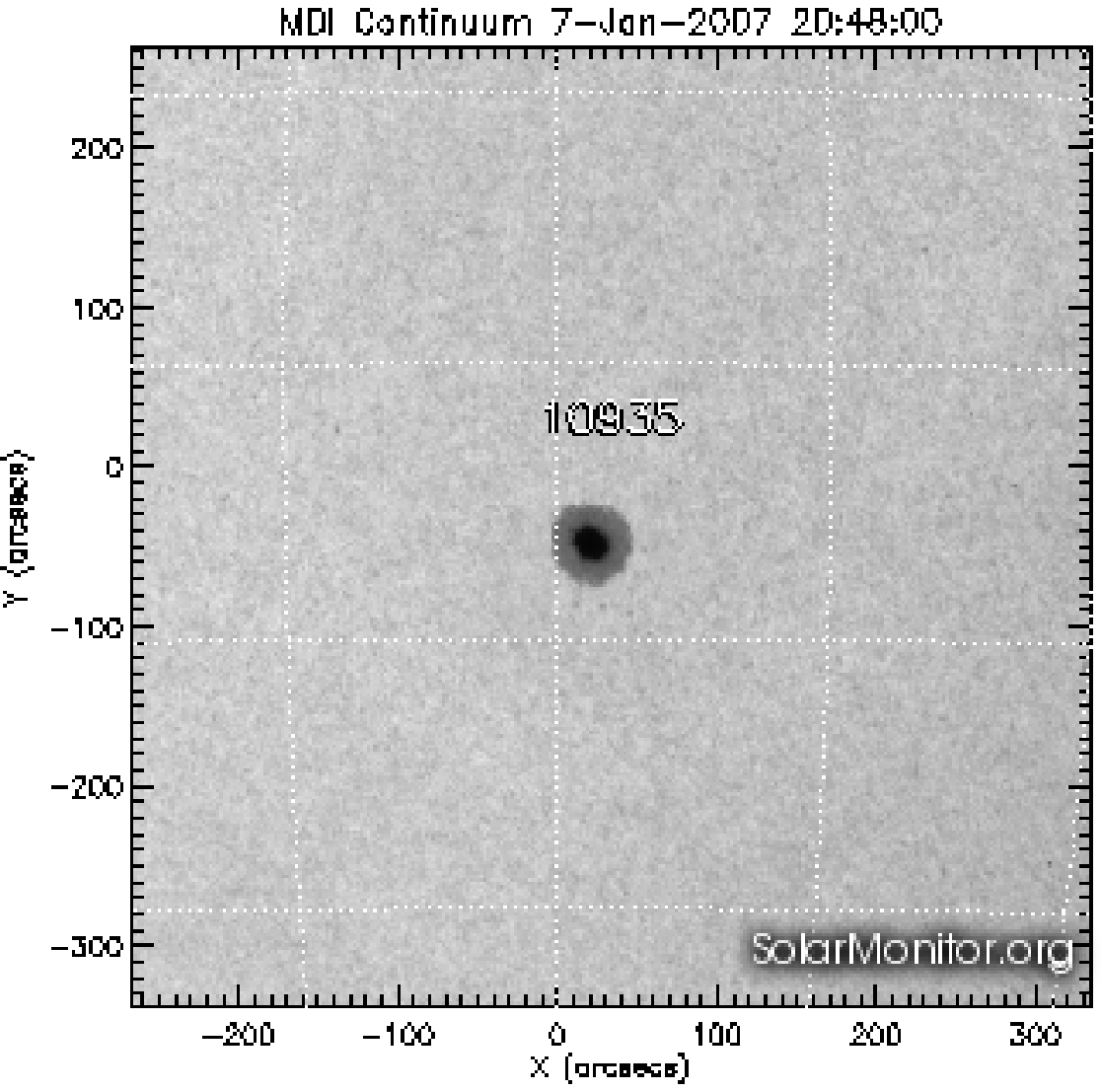}\\
\includegraphics[width=40mm]{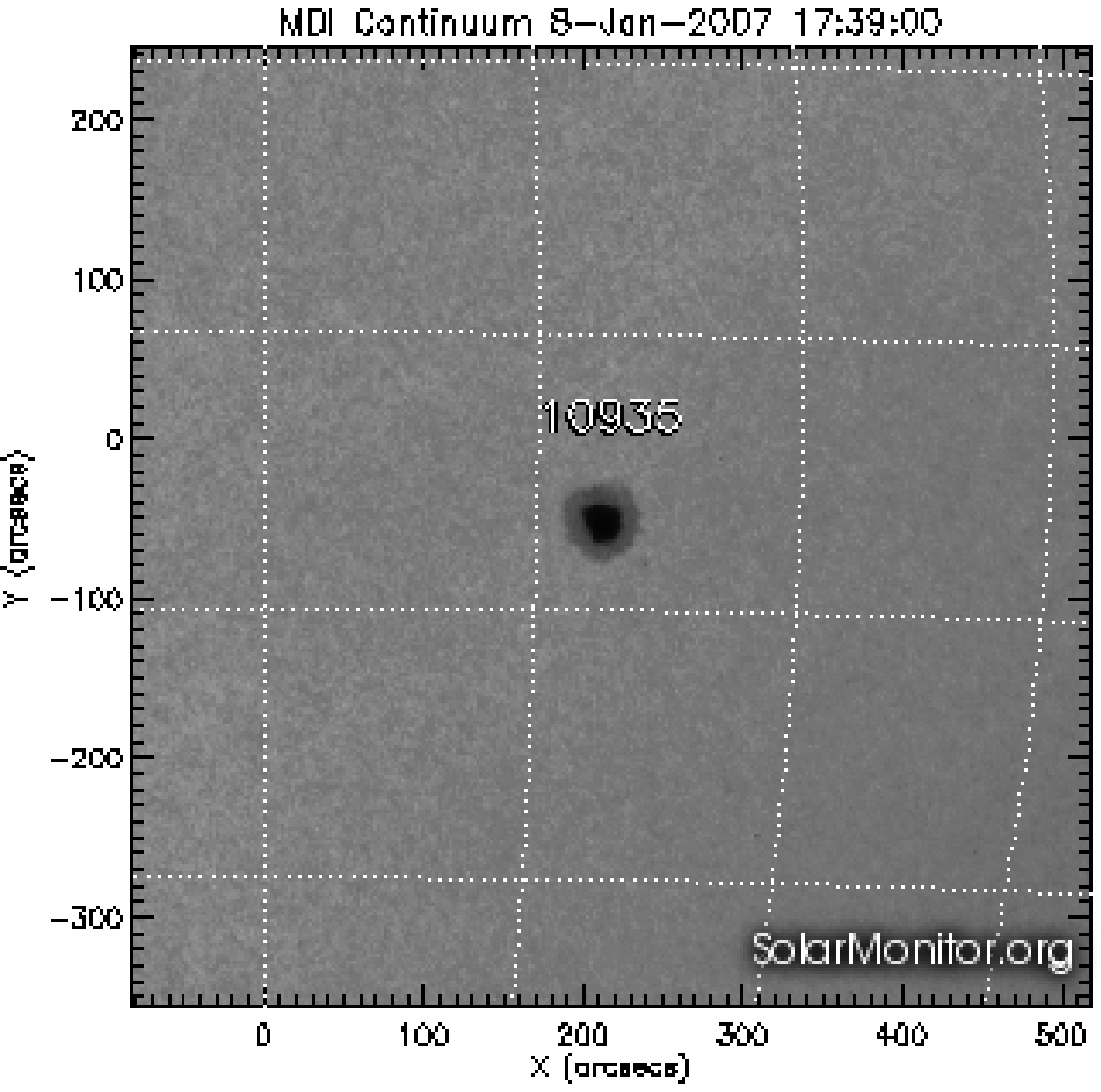}
              }
             \caption{Examples of MDI continuum images of active regions used in this study 
for three consecutive
 days. Top: AR 10923 with non-rotating sunspots; Middle: AR 10930 with a group of non-rotating and 
rotating sunspots; Bottom:
 AR10935 with a non-rotating sunspot. These images are taken from url {\it http://www.solarmonitor.org}.
    }
   \label{fig1}
   \end{figure}

\section{Selection of Data}

In this study, we focus on three isolated active regions: ARs 10923, 10930 and 10935. 
Sample continuum images of these regions for three consecutive days are shown in Figure~1.
All three regions were observed around the same latitude band in the
southern hemisphere in three successive Carrington
 rotations during the declining phase of solar cycle 23 and located in areas relatively free 
from other magnetic activity.  In addition, our analysis is confined to those days when all three regions were also
positioned at the same heliographic locations. This selection minimizes the influence of systematic 
errors on inferences that may arise due to 
the location of regions analyzed. Since new region numbers are assigned whenever regions appear 
on the east limb irrespective of their reappearance after completing a  solar rotation, AR 10930 
appears to be the same as 10923 in the preceding rotation and AR 10935 in the following rotation. 
However, all these regions exhibited different characteristics during their front-side disk passages
and are considered independent active regions.
It can be seen in Figure~1 that AR 10930 had two major sunspots; the big sunspot did 
not show any visible change during the  disk passage  while the small sunspot in the 
southern part exhibited rapid counterclockwise rotation about its umbral center.  This sunspot 
started to rotate after mid day on 10 December 2006, continued to rotate until 13 December 2006
and finally produced a large X-3.4 class flare. However, sunspots in the other two active regions did
not show any evidence of their rotation within their penumbral boundaries, hence ther are refereed to as ``non-rotating''
sunspots. 

\begin{table}
\begin{center}
\caption{Details of active regions analyzed.}
\label{tlab}
\begin{tabular}{cccccccc}\hline
CR & Active & Day   & Start & End &  \multicolumn{2}{c} {Location} & CR \\
   & region & number& Date ~~~Time${^a}$ &  Date  ~~~Time${^a}$  & Long & Lat &Long \\
\hline
2016&10923& &&&&\\
&&1&12-Nov-06 09:01& 13-Nov-06 13:00& -19.00$^{\mathrm{o}}$ & -4$^{\mathrm{o}}$ & 5.9$^{\mathrm{o}}$\\
&&2&13-Nov-06 09:01& 14-Nov-06 13:00& -5.82$^{\mathrm{o}}$ & -4$^{\mathrm{o}}$ & 5.9$^{\mathrm{o}}$\\
&&3&14-Nov-06 09:01& 15-Nov-06 13:00& 7.36$^{\mathrm{o}}$ & -4$^{\mathrm{o}}$ & 5.9$^{\mathrm{o}}$\\
&&4&15-Nov-06 09:01& 16-Nov-06 13:00&  20.54$^{\mathrm{o}}$ & -4$^{\mathrm{o}}$ & 5.9$^{\mathrm{o}}$\\

\\
2017&10930& &&&&\\
&&1&09-Dec-06 10:01& 10-Dec-06 12:00&-19.00$^{\mathrm{o}}$ & -4$^{\mathrm{o}}$ &9.5$^{\mathrm{o}}$\\
&&2&10-Dec-06 10:01& 11-Dec-06 12:00& -5.82$^{\mathrm{o}}$ & -4$^{\mathrm{o}}$ &9.5$^{\mathrm{o}}$\\
&&3&11-Dec-06 10:01& 12-Dec-06 12:00& 7.36$^{\mathrm{o}}$ & -4$^{\mathrm{o}}$ &9.5$^{\mathrm{o}}$\\
&&4&12-Dec-06 10:01& 13-Dec-06 12:00  &20.54$^{\mathrm{o}}$ & -4$^{\mathrm{o}}$ &9.5$^{\mathrm{o}}$\\

\\
2018&10935 & &&&\\
&&1 & 05-Jan-07 17:55& 06-Jan-07 21:54& -19.00$^{\mathrm{o}}$ & -4$^{\mathrm{o}}$ & 9.5$^{\mathrm{o}}$\\
&&2 & 06-Jan-07 17:55& 07-Jan-07 21:54& -5.82$^{\mathrm{o}}$ & -4$^{\mathrm{o}}$ &9.5$^{\mathrm{o}}$\\
&&3 & 07-Jan-07 17:55& 08-Jan-07 21:54& 7.36$^{\mathrm{o}}$& -4$^{\mathrm{o}}$ &9.6$^{\mathrm{o}}$\\
&&4 & 08-Jan-07 17:55& 09-Jan-07 21:54& 20.54$^{\mathrm{o}}$& -4$^{\mathrm{o}}$&9.6$^{\mathrm{o}}$\\

\hline
\end{tabular}
\end{center}
${^a}$ All times are in UT.
\end{table}

Our analysis utilizes high-resolution continuous Doppler data obtained by the  
Global Oscillation Network Group (GONG) and 96-min cadence magnetograms from the MDI
onboard {\it Solar and Heliospheric Observatory} (SOHO).
The start and end times of the various data sets used here are given in Table~1.
While Dopplergrams are used to infer components of horizontal velocity using
the ring-diagram method, the magnetograms
provide estimates of a magnetic activity index (MAI) as a measure of solar activity.
To obtain the MAI, we convert magnetogram data to absolute values, average
 over the length of a ring day and apodise them into circular areas to match  the size of the 
Dopplergram patches used in the ring-diagram analysis.  The temporal variation of MAI in all three active regions 
is plotted in Figure~2. It is seen that while the average magnetic flux for ARs 10923 and 
10930 has an increasing trend with time, it decreases for AR 10935, thus we categorize the former 
two as emerging active regions and the later as decaying. 
Since  we do not have the resolution to discriminate between sunspots of different 
topology within the same active region for a meaningful analysis using ring-diagram technique, 
we will consider AR 10930 as a ``rotating sunspot" and, ARs 10923 and 10935 
as ``non-rotating sunspots".

\begin{figure}   
   \centerline{
\includegraphics[width=100mm]{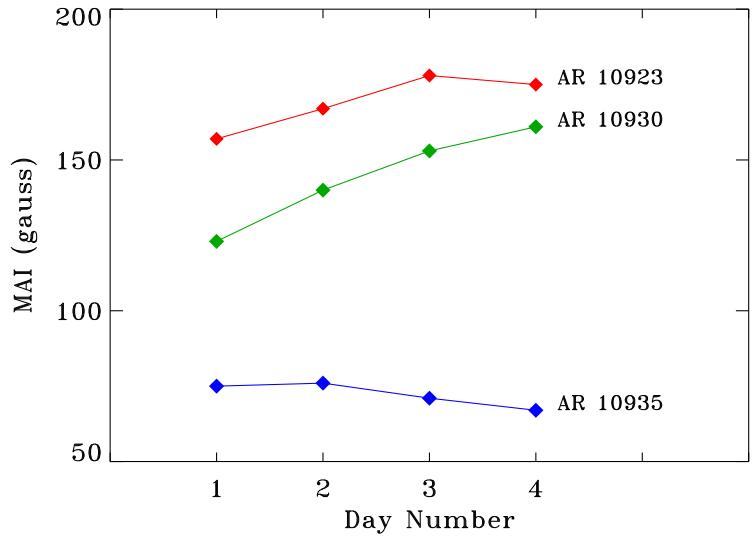}
              }
            \caption{Temporal variation of magnetic activity index (MAI) in three active regions.}
   \label{fig2}
   \end{figure}

\section{Analysis and Results}

In this Section, we derive the subsurface horizontal velocity and its variation with time for 
each active region. It should be noted that areas used here are smaller in size than those used 
in the standard ring-diagram analysis where areas of 16$^{\mathrm{o}}\times$16$^{\mathrm{o}}$ are 
generally analyzed (\opencite{haber00}; \opencite{corbard03} \opencite{komm08}).   The analysis of smaller areas is
particularly useful in the study of active regions where we can  minimize the influence of 
relatively quiet neighboring areas  (\opencite{jain10}, \citeyear{jain11a}). However,  major disadvantages of this 
selection combined with the spatial resolution of images and the limitation of the 
ring-diagram technique include restrictions on i) fitted modes that are available for inversion
and,  ii) the depth range that can be  probed reliably.  In this article, we discuss the relative variation in 
both zonal and meridional components of the horizontal velocity with time at ten target 
depths in the range from surface to about 12 Mm.
The magnitudes are different from those reported in previous articles using ring-diagram technique
({\it e.g.} \opencite{komm11}; and references therein) where residuals, obtained by removing large 
scale flows from inverted velocities, were widely studied.

\begin{figure}
\includegraphics[width=120mm]{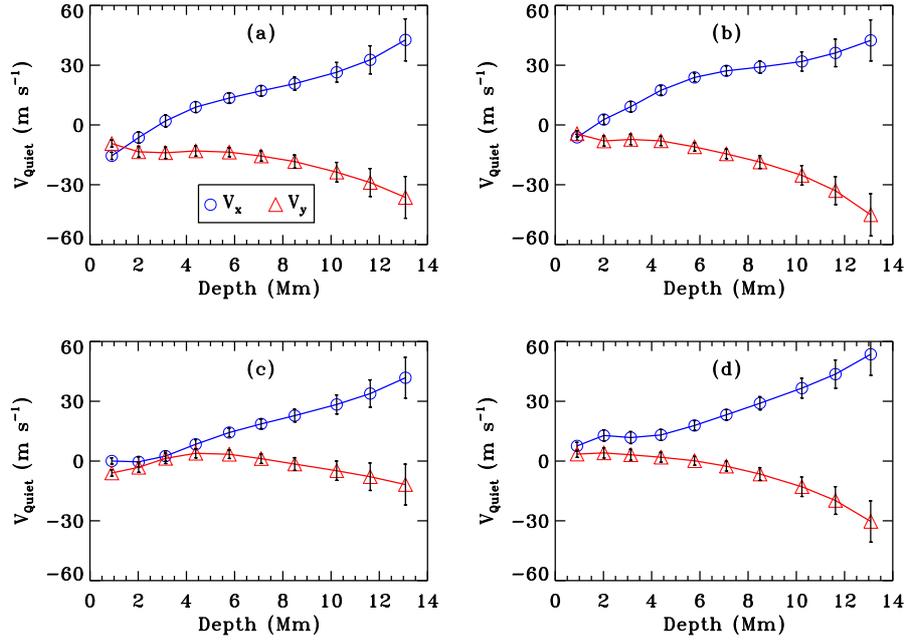}\\
             \caption{Depth variation  of average $V_x$ (blue circles) and $V_y$ (red triangles)
for quiet regions at four different locations along latitude 4$^{\mathrm{o}}$S. The
heliocentric longitudes are: (a) 19$^{\mathrm{o}}$E, (b) 5.82$^{\mathrm{o}}$E, (c) 7.36$^{\mathrm{o}}$W
and (d) 20.54$^{\mathrm{o}}$W.
    }
   \label{fig3}
   \end{figure}

In order to investigate the  temporal evolution of horizontal velocity beneath 
active regions in various time samples, the first step is to eliminate the influence 
of systematic effects, if any. One potential source of error is the effect of
projection that may arise  due to the change in location of tracked 
regions from day to day during the disk passage \cite{komm11}. Since the ring-diagram
technique is sensitive to such effects, it is necessary to ensure that our inferred values are
not affected by such systematics. To overcome this artifact, we calculate reference 
velocities in several quiet regions at the same heliographic locations as for the active regions 
given in Table~1 and their surrounding regions.  We derive best estimates 
of the quiet-Sun velocities by using  error-weighted averages of eight ring days 
with no noticeable magnetic activity  on the visible disk chosen from Carrington 
rotations considered in this analysis. 

Figure~3 shows the variation of average $x$ (zonal) 
and $y$ (meridional) components of the horizontal velocity,  ($V_x$  and  $V_y$),  
as a function of depth for quiet regions at the same four locations as active regions. 
As expected, in all cases, the magnitude 
of both $V_x$ and  $V_y$  increases with depth. This is consistent with observations 
where rotation rate near the equator is found to increase with depth in 
the outer 5\% of the Sun and also the meridional component is poleward. Negative values of
 $V_y$ indicate the south-poleward direction of the meridional component of the velocity.
These profiles will be used to eliminate the contribution of quiet regions from the velocity components
of active regions. As mentioned above, the projection effects in both quiet and active regions 
are expected to be similar at same heliographic locations. Hence, we subtract average velocities 
of  quiet regions from the calculated velocities in and around active regions. This ensures that 
systematic effects as well as zonal and meridional components of quiet regions are removed 
from the active region values which are discussed below.

\begin{figure}   
   \centerline{
\includegraphics[width=120mm]{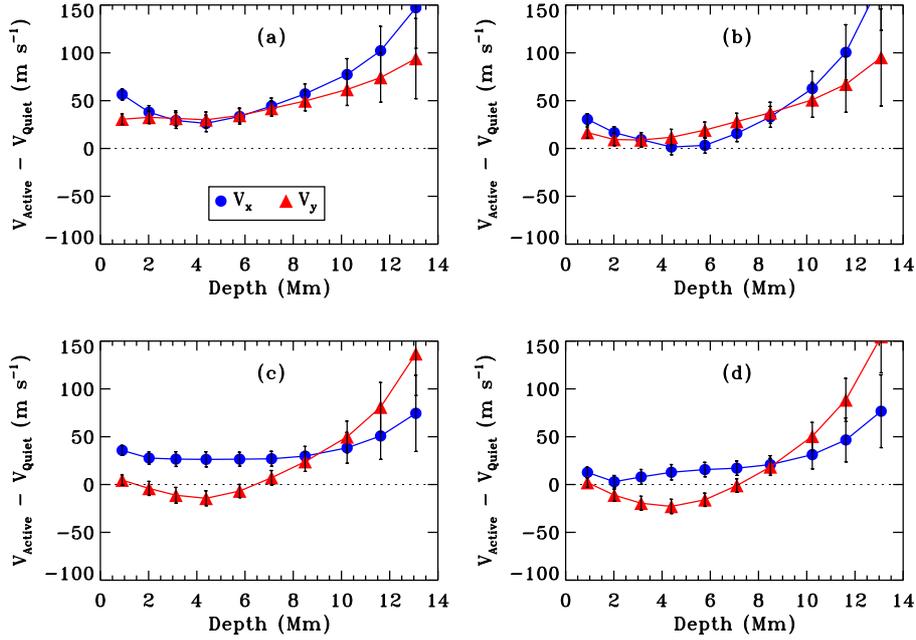}
              }
             \caption{Depth variation of $V_x$ (blue circles) and $V_y$ (red triangles) 
for an emerging active region (AR 10923) consisting of  non-rotating sunspots for four 
different epochs  given in Table~1; 
(a) Day  1, (b)  Day 2, (c) Day 3, and (d) Day 4. Values are plotted at target depths only.
    }
   \label{fig4}
   \end{figure}

 \begin{figure}   
   \centerline{
\includegraphics[width=120mm]{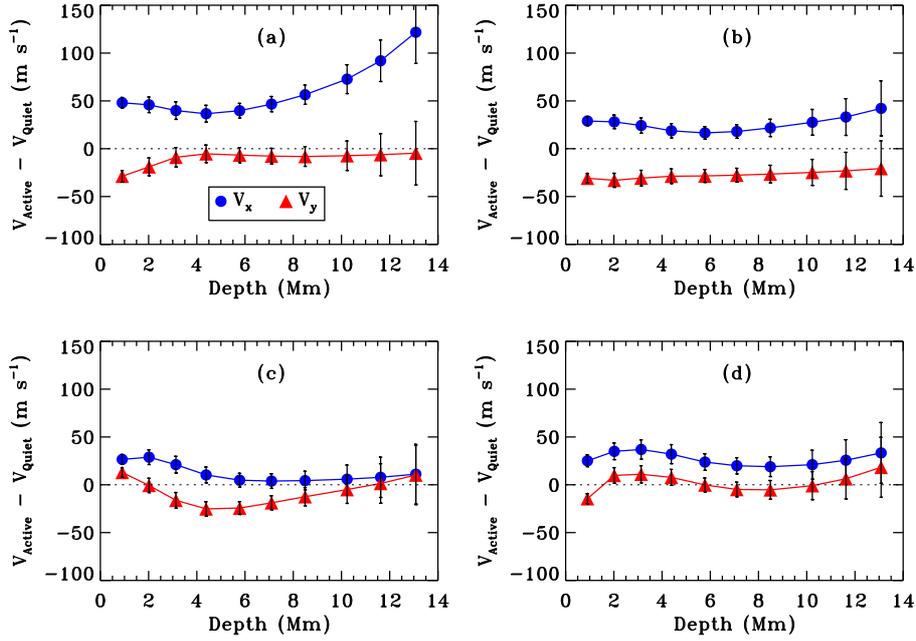}
              }
             \caption{Depth variation of $V_x$ (blue circles) and $V_y$ (red triangles) 
for a decaying active region (AR 10935) consisting of  non-rotating sunspots for four 
epochs  given in Table~1; 
(a) Day  1, (b)  Day 2, (c) Day 3, and (d) Day 4. Values are plotted at target depths only.
    }
   \label{fig5}
   \end{figure}

\subsection{Rotating and Non-rotating Sunspot Regions}

Figures~4 and 5  show the variation of $V_x$  and  $V_y$ with depth for four epochs
for ARs 10923 and 10930, respectively. Both components are seen to exhibit a systematic 
variation with depth.  $V_x$ is predominantly positive in all cases, independent of 
the epoch, which clearly shows that the zonal component in active regions is faster 
than in the quiet region. Comparing Figures 4 and 5, we find much higher  $V_x$  in the 
emerging region AR 10923  than the decaying AR 10935. In addition, the MAI for AR 
10923 is significantly higher with rapidly increasing trend for the first few  days 
while this variation is opposite for AR 10935.  Thus, the zonal component in the strong
emerging-flux region exhibits considerable variation with depth while it is comparable in the 
decaying-flux region to that of quiet regions. These findings support the previous analysis 
({\it e.g.} \opencite{komm11}) where average flows for a large number of emerging and decaying 
active regions were studied. In contrast, the meridional component exhibits a different
 behavior.  It increases sharply in deeper layers for the emerging-flux region while it
remains mostly unchanged in the case of the decaying active region.

\begin{figure}
\includegraphics[width=120mm]{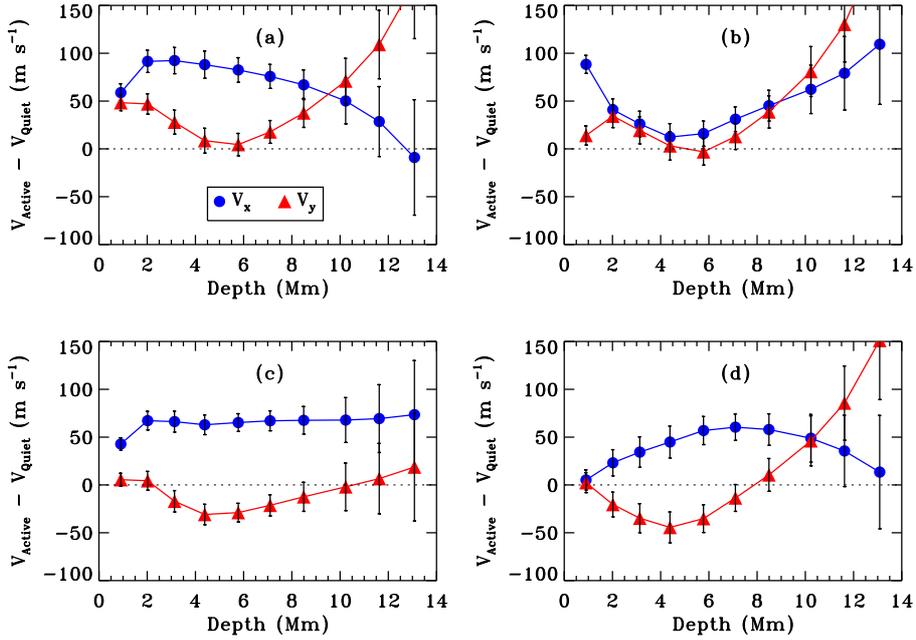}\\
             \caption{Depth variation of $V_x$ (blue circles) and $V_y$ (red triangles) 
for an active region (AR 10930) consisting of  rotating and non-rotating sunspots for four 
epochs  given in Table~1; 
(a) Day  1, (b)  Day 2, (c) Day 3, and (d) Day 4. Values are plotted at target depths only. 
    }
   \label{fig6}
   \end{figure}

 In Figure~6, we display similar plots for AR 10930 where one of the sunspots rotates around 
its umbral center. In contrast to other two active regions, the zonal component associated
with this active region varies 
significantly with depth from one day to another. Although it remains positive for 
all ring days, similar to other active regions, the depth profile changes abruptly. Also, while the depth
 profile  of the meridional component is similar for all four time samples, these exhibit significant
variations in magnitude. To understand these observations, we focus on the depth range between 
2 Mm and 12 Mm, where the errors are relatively small due to the set of modes contained in the analysis
 combined with the limitations of the model used near the surface. Below 2 Mm, we find that  
$V_x$, prior to the sunspot rotation (Figure~6(a)),  decreases
 rapidly in deeper layers while its magnitude  in Figure~6(b) first decreases up to 5~-~6 Mm and then 
increases. Note that Figure~6(b) includes the initial period of sunspot rotation.
Further, $V_x$ does not change significantly in Figure~6(c) and finally, 
 it shows a trend opposite to that was seen in Figure~6(b).

\begin{figure}
\includegraphics[width=120mm]{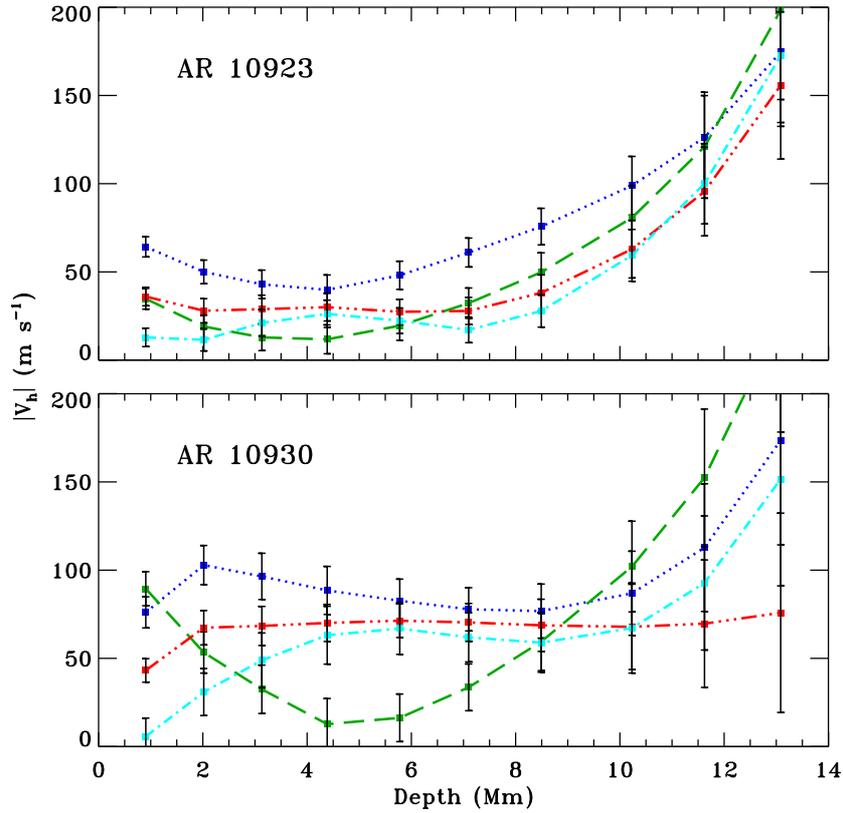}\\
             \caption{Variation of total horizontal flow as a function of depth for 
 emerging flux regions AR 10923 (upper panel) and 10930 (lower panel) for four
overlapping time intervals given in Table~1; (blue dotted) Day 1, (green dashed) Day 2, (red dashed-dot-dot-dot)
Day 3, and (cyan dashed-dot) Day 4.  Values are plotted at target depths only.
    }
   \label{fig7}
   \end{figure}

To interpret the physical origin of the sudden changes in the depth profiles of AR 10930 with time, 
we compare its inferred velocities with those of AR 10923. Although, in both cases, 
the magnetic flux is increasing with time, there
is a significant difference in the morphology of these regions; AR 10930 has a sunspot  
rapidly rotating in the counter-clockwise direction  while all sunspots in AR 10923 
are non-rotating. Despite the temporal variation in MAI for AR 10923, its depth profile, 
in the absence of sunspot rotation, does not show any significant variation from one day 
to another while the rotation in AR 10930 is associated with changes  for different time samples. This can be easily 
seen in Figure~7 where we plot the total horizontal velocity, defined as the square root of the sum
of the squares of the velocity components. AR 10923 has
smooth and similar variations with depth for all four days, while AR 10930 displays significant 
variation. The maximum change is seen at the beginning of sunspot rotation (Day 2)
where total velocity decreases up to 4--5 Mm and then increases sharply in deeper layers providing 
an evidence for flows in two different directions. This is in agreement with  \inlinecite{zhao03}
who report opposite flows in a rotating sunspot group (AR 9114),
 but this feature is absent in a non-rotating sunspot group. 

\subsection{Surrounding Regions}

  \begin{figure}   
   \centerline{\includegraphics[width=120mm,]{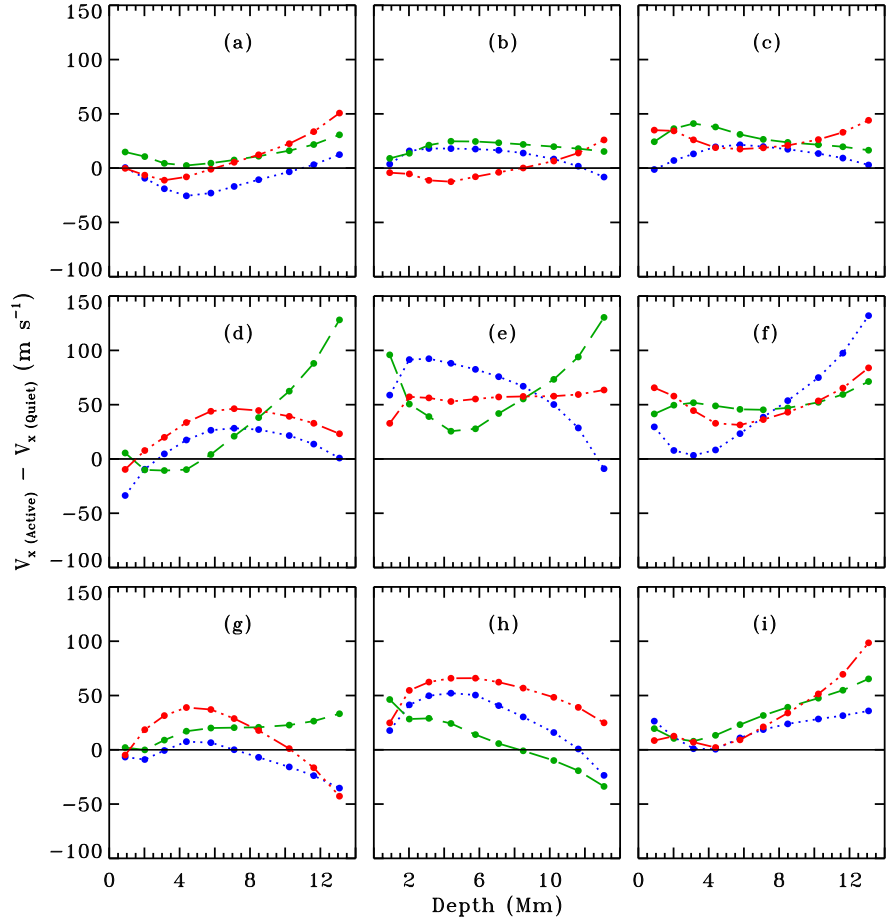}
              }
              \caption{Depth variation of $V_x$  in and around AR 10930 for 
three overlapping time series; blue dotted, green dashed and red dashed-dot-dot  lines
represent variation for Days 1, 2, and 3, respectively.  Errors are of 
the same order as in Figure~6. The active region is located at the center 
 of Panel (e).   }
  \label{fig8}
   \end{figure}

 \begin{figure}  
  \centerline{\includegraphics[width=120mm,]{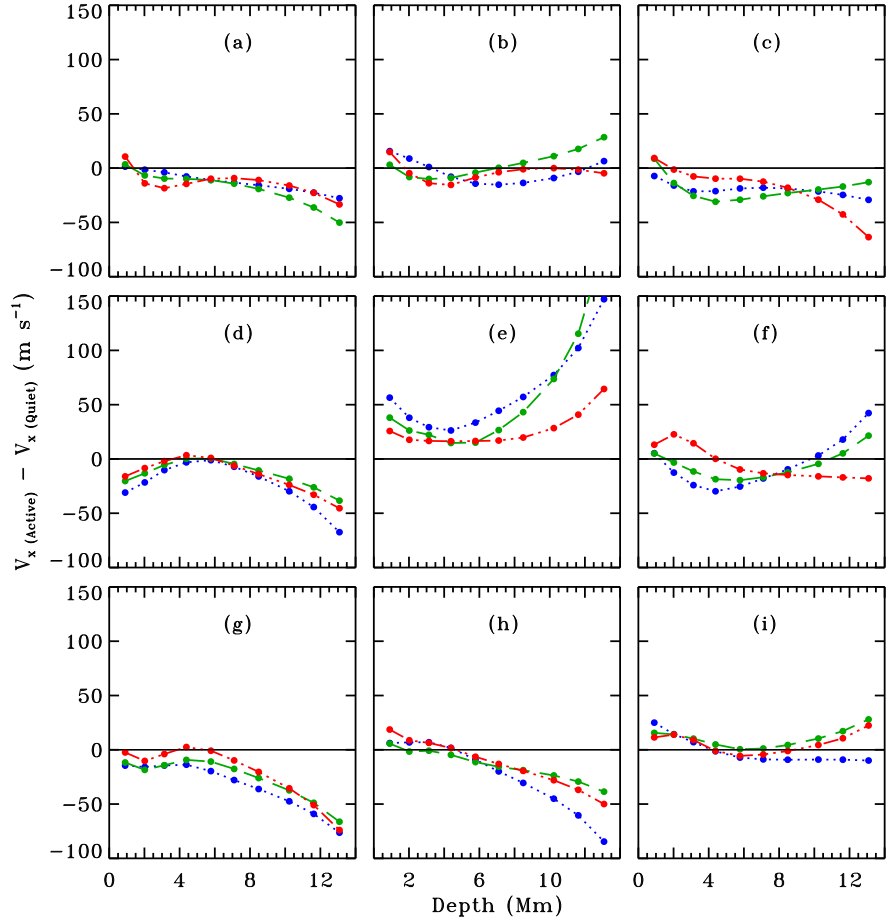}
              }
              \caption{Depth variation of $V_x$  in and around AR 10923 for 
three overlapping time series; blue dotted, green dashed and red dashed-dot-dot  lines
represent variation for Days 1, 2, and 3, respectively.  Errors are of 
the same order as in Figure~4. The active region is located at the center 
 of Panel (e). }
   \label{fig9}
   \end{figure}

It has been shown earlier that sub-surface flows in areas containing active regions 
are larger than their quieter surroundings \cite{komm05a}. Here we extend this analysis 
to quiet areas surrounding the rotating and non-rotating sunspots and analyze a mosaic of nine
 regions for three consecutive ring days. The central region in
each mosaic has the complete active region at its center ({Panel (e) of Figures~8 through 11}) while the neighboring eight regions, spaced by 5$^{\mathrm{o}}$ in each direction from the central region, are relatively quiet as they include only a part of the  active region. The  results for the temporal variation of zonal and meridional components around ARs
10930 and 10923 are shown in Figures~8 through 11;  Figures~8 and 9 focus on the variation of the zonal components, while Figures~10 and 11 provide information on the meridional components. In general, it is observed 
that both the zonal and meridional components of the central region (Panel (e) of these figures) have the maximum variation with depth while the neighboring quiet regions with smaller magnetic flux display smaller variation.

  \begin{figure}   
   \centerline{\includegraphics[width=120mm,]{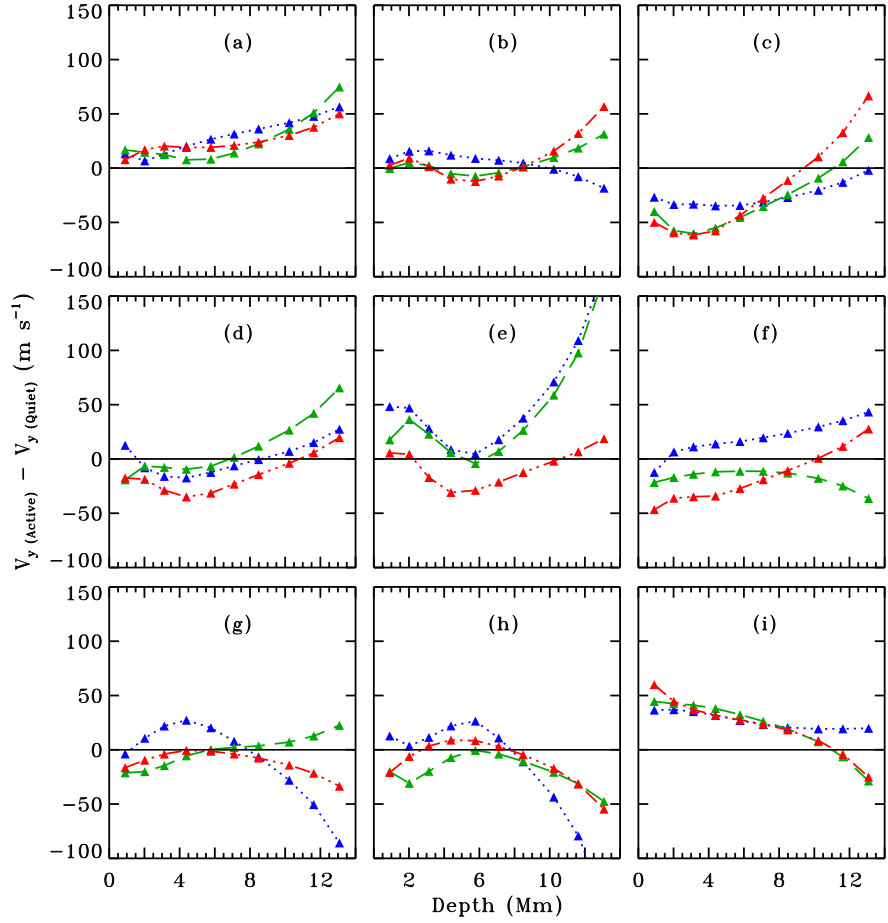}
              }
              \caption{Depth variation of $V_y$ in and around AR 10930 for 
three overlapping time series; blue dotted, green dashed and red dashed-dot-dot lines
represent variation for Days 1, 2, and 3, respectively.  Errors are of 
the same order as in Figure~6. The active region is located at the center 
 of Panel (e).   }
   \label{fig10}
   \end{figure}

 \begin{figure}  
  \centerline{\includegraphics[width=120mm,]{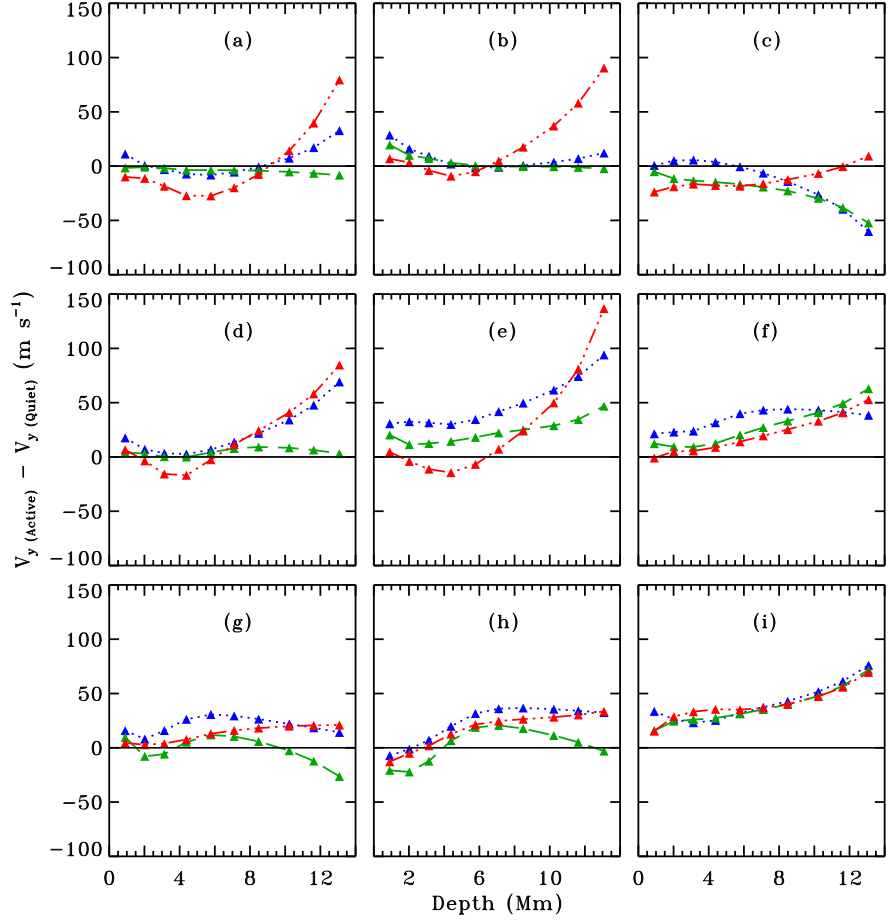}
              }
              \caption{Depth variation of $V_y$ in and around AR 10923 for 
three overlapping time series; blue dotted, green dashed and red dashed-dot-dot lines
represent variation for Days 1, 2, and 3, respectively.  Errors are of 
the same order as in Figure~4. The active region is located at the center  
 of Panel (e).  }
   \label{fig11}
   \end{figure}

On a detailed comparison of  $V_x$ of neighboring regions, we note that the depth profiles 
of AR 10923 for each day are comparable (within 1$\sigma$ estimates) while these 
 are different in a few cases for AR 10930.  
In particular, the profiles of $V_x$ in the regions that are adjacent to the rotating part
of the AR 10930  exhibit reasonable temporal variations 
(shown in Panels (d), (g), and (h)) and suggest that the variation is induced by the 
rotation of the sunspot.  Similar results are 
obtained for meridional components (Figures~10 and 11). This provides
evidence that a rotating sunspot not only changes the flow pattern in the active region
but also the flows in the surrounding areas. 

\section{Summary}
We have studied the horizontal velocity in a limited number of active regions consisting 
of rotating and non-rotating sunspots. The analysis presented here is based on three 
active regions ARs 10923, 10930, 10935 from three consecutive Carrington rotations: 
AR 10930 contains a fast-rotating sunspot in a strong emerging active region 
while other two have non-rotating sunspots with emerging flux in AR 10923 and decaying 
flux in AR 10935. In agreement with earlier results, we note that active regions have 
faster zonal component as compared with the quiet regions. The variations are also 
found to be higher in emerging-flux regions as compared to the decaying regions. A 
comparison between the plasma flow between regions with rotating and non-rotating 
sunspots illustrates that the flow associated with the rotating sunspot in AR 10930 
varies significantly with depth during the course of the rotation  while the variations 
in the flow in non-rotating sunspot groups, ARs 10923 and 10935, are primarily 
constant (within 1$\sigma$). We further find evidence of two opposite flows at 
different depth  and conjecture that this opposite flow provides a twist in magnetic 
field lines. We illustrate that the flows in areas surrounding the active region  
in both cases are relatively small, but there is a significant variation in regions 
neighboring to the rotating sunspot.

In the present study, a sunspot group consisting of both rotating and non-rotating 
sunspots is considered as a rotating sunpots as a whole,
thus a complete picture of the sub-surface flows beneath a rotating sunspot
could not be obtained. This is mainly due to the limitations imposed by the spatial resolution 
 on the technique. Although we have found significant differences between the
temporal variations in depth profiles of rotating and non-rotating sunspots, 
the analysis of smaller areas will further provide a deeper insight into the dynamics of
sub-surface layers below individual sunspots.  We plan to carry out a  detailed
study using  images with improved spatial resolution from the  Helioseismic
 Magnetic Imager  (HMI) onboard {\it Solar Dynamics Observatory} (SDO)  that will 
allow us to analyze smaller regions within an active region.  Our preliminary
analysis using HMI Dopplergrams clearly shows that the ring-diagram technique can be 
reliably applied to regions as small as 5$^{\mathrm{o}}$  \cite{jain11b} and one
can obtain a comprehensive picture  of the 
solar sub-surface layers which is crucial for space weather studies.

\begin{acks}
This work utilizes data obtained by the Global Oscillation Network
Group (GONG) project, managed by the National Solar Observatory, which
is operated by AURA, Inc. under a cooperative agreement with the
National Science Foundation. The data were acquired by instruments
operated by the Big Bear Solar Observatory, High Altitude Observatory,
Learmonth Solar Observatory, Udaipur Solar Observatory, Instituto de
Astrof\'{\i}sica de Canarias, and Cerro Tololo Interamerican
Observatory. It also utilizes data from the Solar Oscillations Investigation/Michelson Doppler Imager
 on the Solar and Heliospheric Observatory. SOHO is a mission of international cooperation
 between ESA and NASA. This work was partially supported by NASA grant NNG08EI54I 
and NSF Award 1062054  to the National Solar Observatory.
\end{acks}

\bibliographystyle{spr-mp-sola}
\bibliography{Jain_rss_May9}

\end{article} 

\end{document}